\documentclass[showpacs,nofootinbib,preprintnumbers,prd,superscriptaddress,twocolumn]{revtex4}

\usepackage{amsmath}
\usepackage{amsfonts}
\usepackage{amssymb}
\usepackage{graphicx}
\usepackage[titletoc]{appendix}
\usepackage{color}
\usepackage{hyperref}
\usepackage{cleveref}
\usepackage[rightcaption]{sidecap}
\usepackage{subfigure}
\usepackage{comment}

\usepackage{mathtools}

\usepackage{dcolumn}

\usepackage{array}
\usepackage{ctable}
\usepackage{multirow}
\usepackage{siunitx}
\usepackage{longtable}
\usepackage{tabularx}
\usepackage{booktabs}

\graphicspath{{Graphics/}}

\def\be{\begin{equation}}
\def\ee{\end{equation}}
\def\bea{\begin{eqnarray}}
\def\eea{\end{eqnarray}}

\definecolor{vividviolet}{rgb}{0.62, 0.0, 1.0}
\definecolor{amaranth}{rgb}{0.9, 0.17, 0.31}
\definecolor{palatinateblue}{rgb}{0.15, 0.23, 0.89}
\definecolor{brightpink}{rgb}{1.0, 0.0, 0.5}
\definecolor{cornflowerblue}{rgb}{0.39, 0.58, 0.93}
\definecolor{deepcarminepink}{rgb}{0.94, 0.19, 0.22}
\definecolor{radicalred}{rgb}{1.0, 0.21, 0.37}
\hypersetup{ linktoc=all,
    colorlinks, linkcolor={palatinateblue},
    citecolor={brightpink}, urlcolor={amaranth}
}

\begin{document}

\title{Inflationary entanglement}

\author{Alessio Belfiglio}
\email{alessio.belfiglio@unicam.it}
\affiliation{School of Science and Technology, University of Camerino, Via Madonna delle Carceri, Camerino, 62032, Italy.}
\affiliation{Istituto Nazionale di Fisica Nucleare (INFN), Sezione di Perugia, Perugia, 06123, Italy.}

\author{Orlando Luongo}
\email{orlando.luongo@unicam.it}
\affiliation{School of Science and Technology, University of Camerino, Via Madonna delle Carceri, Camerino, 62032, Italy.}
\affiliation{Istituto Nazionale di Fisica Nucleare (INFN), Sezione di Perugia, Perugia, 06123, Italy.}
\affiliation{Institute of Experimental and Theoretical Physics, Al-Farabi Kazakh National University, Almaty 050040, Kazakhstan.}

\author{Stefano Mancini}
\email{stefano.mancini@unicam.it}
\affiliation{School of Science and Technology, University of Camerino, Via Madonna delle Carceri, Camerino, 62032, Italy.}
\affiliation{Istituto Nazionale di Fisica Nucleare (INFN), Sezione di Perugia, Perugia, 06123, Italy.}

\begin{abstract}
We investigate the entanglement due to geometric corrections in particle creation during inflation. To do so, we propose a single-field inflationary scenario, nonminimally coupled to the scalar curvature of spacetime. We require particle production to be purely geometric, setting to zero the Bogolubov coefficients and computing the $S$ matrix associated to spacetime perturbations, which are traced back to inflaton fluctuations. The corresponding particle density leads to a nonzero entanglement entropy whose effects are investigated at primordial time of Universe evolution. The possibility of modeling our particle candidate in terms of dark matter is discussed. The classical back-reaction of inhomogeneities on the homogeneous dynamical background degrees of freedom is also studied and quantified in the slow-roll regime.
\end{abstract}

\pacs{03.67.Bg, 04.62.+v, 98.80.-k, 98.80.Cq}

\date{\today}
\maketitle

%%%%%%%%%%%%%%%%%%%%%%%%%%%%%%%%%%%%%%%%%%%%%%%%%%%%%%%%%%%%%%%%%%%%%%%%%%

\section{Introduction} \label{sez1}

Throughout the Universe evolution, inflation is realized at primordial epoch with the aim of healing the main issues related to the standard \emph{Big Bang paradigm} \cite{cosmo1,cosmo2}. It represents a phase of strong acceleration, slightly similar to late time \emph{dark energy}\footnote{The cosmography of inflation is similar to dark energy \cite{cosmography1,cosmography2}, but physically quite different. Models,  unifying the two scenarios, are however object of current investigation \cite{unione}.} epoch \cite{cosmo3}, driven by an \emph{inflaton field}\footnote{The idea of considering generic scalar fields deliberately represents  the simplest case, describing the inflaton. Alternatives, not fully-excluded by observations, comprehend Higgs field, spinor fields, etc., see e.g. \cite{infl1,infl2, ref5}. }, quite different of barotropic fluids, widely adopted for established dark energy scenarios. The inflationary potential is still unknown and can be built up using the approach of either \emph{small} or \emph{large} fields, with exceptions provided by \emph{intermediate field representations} \cite{cosmo4}, or by means of couplings among more than one field, see e.g. \cite{more}.

One of the main goals of inflation is to reproduce \emph{inhomogeneities} responsible for the formation of large-scale structures \cite{strutture}. Thus, inflation appears to be the natural landscape in which overdensities formed at primordial time. For any inflationary potentials, as a consequence of Einstein's field equations, the cosmological inhomogeneities plausibly generate particles, a mechanism known as \emph{geometric particle production} \cite{ref1,ref2}. Such process is conceptually different from the well known gravitational particle production (GPP) from vacuum, which is typically associated to Bogolubov transformations for quantum fields in an expanding unperturbed spacetime \cite{gp1,gp2,gp3,gp4, gp5, lyth}.

Indeed, assuming a homogeneous and isotropic Friedmann-Robertson-Walker (FRW) background\footnote{Along the text, we  only focus on the spatially flat version of the FRW spacetime, in agreement with current measurements indicating it as the most accredited scenario, see e.g. \cite{planck}.}, GPP leads to particle
pairs with opposite momenta, but including inhomogeneities, i.e., departing from a genuine FRW, may lead to pair-creation probability depending on local geometric quantities\footnote{We clearly expect geometry to depend on the details of the expansion, so that the two mechanisms are not completely independent from each other, as we will discuss in the manuscript.}. Both mechanisms were shown to produce also entanglement entropy \cite{ball,ent1,ent2,ent3,ent4,ent5,luo,bel1}, and the topic of \emph{cosmological quantum entanglement} has attracted great attention in recent years \cite{entclm}. In fact, quantum correlations arising from particle creation may contain information about the Universe expansion and in principle entanglement could also be extracted directly from spacetime itself \cite{harv1,harv2}. At the same time, most predictions of quantum field theory are indeed difficult to test directly, paving then the way for some analogue models, see e.g. \cite{an1,an2,an3}.

One important motivation in studying particles from inhomogeneities during inflation is due to the fact that such mechanism may be responsible for dark matter production at early times \cite{bel2}, under the usual assumption that the corresponding dark matter candidate is coupled only to gravity and not to other quantum fields\footnote{A similar approach for dark matter production has been recently studied also in the context of GPP \cite{add1,dm1,dm2,dm3,dm4,dm5}, with generalizations to nonzero spin \cite{dm6,dm7,dm8,dm9,dm10,dm11,dm12}.}. Accordingly, if dark matter has negligible interactions with standard matter, quantum correlations created at early times may still be present to some extent at late times, since decoherence due to coupling with other fields (except for gravity) would be excluded. So, if spacetime geometry affects entanglement, a given perturbed FRW background induced by inflationary dynamics is expected to work analogously, leading to non-negligible effects.

Motivated by these considerations, we here investigate entanglement arising from geometric particle production in a single-field inflationary scenario\footnote{Multifield inflation may also lead to interesting results in the context of geometric particle production, starting for example from  the proposal of Ref. \cite{tsu}. This could be subject of future investigations.}, where perturbations are traced back to quantum fluctuations of the scalar inflaton field. By assumption, the inflaton dominates the energy density of the Universe during inflation. Accordingly, any fluctuation in the inflaton results, through Einstein's equations, in a perturbation of the metric. The dynamics of these fluctuations will be then responsible for the geometric mechanism of particle production here studied. In addition, we consider from a classical perspective the back-reaction effects induced by inhomogeneities on the homogeneous dynamical background \cite{abr1,abr2,revbac}.  We start by assuming in our Lagrangian a Yukawa-like term, i.e., a non-minimal interacting term between the inflaton and the scalar curvature. The Universe evolution during inflation is modeled by a quasi-de Sitter solution \cite{ref4} for the scale factor, in the perturbed FRW background. We evaluate then the modes and the corresponding analytical solutions for the inflaton field involving a chaotic potential. Once obtained the e-folding number, the perturbation solution and the end of inflation, we go further with particle production up to the second geometric order, taking zero Bogolubov coefficients at first order expansion. The corresponding geometric particles are thus computed together with their probabilities for positive and negative coupling constant, $\xi$. We infer the amplitude element, adopting the Dyson expansion over the $S$ matrix and afterwards we focus on back-reaction effects. As a final step, the entanglement entropy is computed, showing how it increases in case of negative coupling constant, $\xi$. Physical consequences on inflationary dynamics, dark matter abundance under the form of geometric particles and entanglement signature are also debated.

The paper is structured as follows. In Sec. \ref{sezione2} we work out our cosmological framework, introducing the corresponding single-field description. In Sec. \ref{sez2}, quantum fluctuations are investigated by perturbing the field and the FRW metric. Afterwards, in Sec. \ref{sez3}, inflation is studied as one adopts a quasi-de Sitter scale factor, getting rise to perturbed solutions for the field itself. Once evaluated the e-folding number and the corresponding inflationary end, we shift to particle production in Sec. \ref{sez4}, where we also compare our geometric mechanism of production to inflationary particle production in warm inflation scenarios. In Sec. \ref{sez5}, we investigate how classical back-reaction effects occur in the primordial Universe, emphasizing that they slightly contribute to the overall shift of the energy-momentum tensor, thus being negligible in our framework. Finally, entanglement due to geometric production is quantified in Sec. \ref{sez6}. Conclusions and perspectives are discussed in Sec. \ref{conclusioni}\footnote{Throughout the paper, we adopt natural units, i.e., $\hbar=c=1$, while the metric tensor is taken with signature $(+---)$.}.

\section{Cosmological setup of inflationary dynamics}\label{sezione2}

We start from the usual Lagrangian density for the inflaton $\phi$, introducing a finite coupling $\xi$ between the field itself and the scalar curvature $R$,

\be \label{Lagr}
\mathcal{L}= \frac{1}{2} \left[ g^{\mu \nu} \phi_{, \mu} \phi_{, \nu}- \xi R \phi^2 \right] - V(\phi).
\ee
The potential $V(\phi)$ is left unspecified for the moment, whereas the FRW line element, in cosmic time $t$, reads
\begin{equation}
ds^2= dt^2-a^2(t) d{\bf x}^2\,.
\end{equation}
Thus, we take the variation of the action for Eq. \eqref{Lagr}  with respect to $\phi$, obtaining the equation of motion
\be \label{eqmot}
\ddot{\phi}+ 3H \dot{\phi} - \frac{\nabla^2 \phi}{a^2}+ 6 \xi \left(\dot{H}+2H^2 \right) \phi + V_{, \phi} =0,
\ee
corresponding to the inflaton dynamics, when the FRW background is not perturbed. Here, dot indicates derivative with respect to $t$. In Eq. \eqref{eqmot} the curvature $R$ is function of the Hubble parameter $H = \dot{a}(t)/a$ and $V_{, \phi} \equiv \partial V (\phi)/\partial \phi$. Notice that here the inflaton is still depending on the event $x^\mu\equiv(t,x)$, instead of time coordinate only. In the next section we will split the field $\phi$ in a background homogeneous contribution and quantum fluctuations associated with it.

The dynamics of the inflaton field is more easily evaluated in conformal time, i.e., $\tau = \int dt/a(t)$, where the unperturbed metric tensor becomes
\be \label{confmetric}
g_{\mu \nu}= a^2(\tau) \eta_{\mu \nu},
\ee
namely proportional to the Minkowski metric tensor, $\eta_{\mu\nu}$.
The zero-order equation of motion for the inflaton is then \cite{ref3,ref4}
\be \label{eqpert}
\frac{1}{\sqrt{-g}} \partial_{\mu} \left( \sqrt{-g} g^{\mu \nu} \partial_{\nu} \phi \right) +  \frac{6 \xi \ a^{\prime \prime}}{a^3} \phi+ V_{, \phi}=0,
\ee
where the prime denotes derivatives with respect to conformal time and we made explicit the zero-order scalar curvature \cite{ref2}.
Introducing the \emph{effective potential} for a generic scalar curvature $R$,
\be \label{effectivepot}
V^{\rm eff}(\phi,R)\equiv V(\phi)+ \frac{1}{2} \xi R \phi^2\,,
\ee
that corresponds to an interacting term, non-minimally coupled to curvature, we can therefore rewrite Eq. \eqref{eqpert} as
\be \label{eqpert1}
\frac{1}{\sqrt{-g}} \partial_{\mu} \left( \sqrt{-g} g^{\mu \nu} \partial_{\nu} \phi \right) + V^{\rm eff}_{, \phi }=0\,,
\ee
that holds for any metric tensor $g_{\mu \nu}$.

%%%%%%%%%%%%%%%%%%%%%%%%%%%%%%%%%%%%%%%%%%%%%%%%%%%%%%%%%%%%%%%%%%%%%%%%

\section{Quantum fluctuations during inflation} \label{sez2}

We here introduce perturbations in the aforementioned framework. To do so, we first split the inflaton field as a homogeneous background contribution, say $\phi_0$, plus a term associated to its quantum fluctuations, namely
\be \label{split}
\phi({\bf x},t)= \phi_0(t) + \delta \phi({\bf x},t).
\ee
Second, we employ metric perturbations, whose most general expression for the  line element, describing scalar degrees of freedom, is \cite{ref4,ref5}

\begin{align} \label{scalpert}
ds^2=a^2(\tau) \big[&(1+2\Phi)d\tau^2-2\  \partial_i B\  d\tau dx^i \notag \\
&- \left((1-2\Psi) \delta_{ij}+ D_{ij} E \right) dx^i dx^j \big],
\end{align}
where $\Phi$, $\Psi$, $B$ and $E$ are scalar quantities which depend on space and time coordinates and $D_{ij}\equiv  \partial_i \partial_j- \frac{1}{3} \delta_{ij} \nabla^2$.

Now, the variation of Eq. \eqref{eqpert1} consists in the sum of four different contributions, corresponding to the variations of $\frac{1}{\sqrt{-g}}$, $\sqrt{-g}$, $g^{\mu \nu}$ and $\phi$. By adopting the well-known identity
\be \label{varmet}
\delta g = g g^{\mu \nu} \delta g_{\mu \nu},
\ee
and recalling the zero-th order equation of motion for the field\footnote{From now on, for simplicity we neglect the subscript $0$ for all background quantities, so that we will regard $\phi$ as $\phi_0$.}
\be \label{zeroord}
\phi^{\prime \prime}+ 2 \frac{a^\prime}{a} \phi^\prime= -V^{\rm eff}_{, \phi } a^2,
\ee
one arrives at the first-order perturbed equation
\begin{align} \label{firstord}
&\delta \phi^{\prime \prime}+ 2 \mathcal{H} \delta \phi^\prime- \partial_i \partial^i \delta \phi- \Phi^\prime \phi^\prime - 3 \Psi^\prime \phi^\prime - \partial_i \partial^i B\  \phi^\prime \notag \\
&=  - \xi\delta R\  \phi\  a^2- \left( V^{\rm eff}_{, \phi \phi}\delta \phi\  + 2 \Phi V^{\rm eff}_{, \phi }\right)\ a^2,
\end{align}
where $\mathcal{H} \equiv a^\prime/a$ and the variation of the scalar curvature is \cite{ref4}
\begin{align} \label{varcurv}
\delta R= \frac{1}{a^2} \bigg( &-6 \mathcal{H}\  \partial_i\partial^i B-2 \partial_i \partial^i B^\prime-2 \partial_i \partial^i \Phi-6\Psi^{\prime \prime}-6 \mathcal{H} \Phi^\prime \notag \\
&-18 \mathcal{H} \Psi^\prime-12 \frac{a^{\prime \prime}}{a} \Phi +4 \partial_i \partial^i \Psi+ \partial_k \partial^i D^k_iE  \bigg).
\end{align}

When perturbations are generated by a single scalar field, it can be shown that the perturbation potentials are equal, i.e., $\Phi=\Psi$.
Moreover, choosing the \emph{longitudinal gauge}\footnote{Geometric particle production can be also studied in the \emph{synchronous gauge}, specified by the condition $h_{0 \nu}=0$ \cite{ref2}. In Appendix \ref{appA} we discuss scalar perturbations in this gauge, starting from the potential $\Psi$ derived here.}, namely $E=B=0$, and assuming plane-wave perturbations \cite{ref4,ref6}, i.e., adopting the following ansatz:
\be \label{ansatz}
\delta \phi({\bf x}, \tau)= \delta \phi_k(\tau)\  e^{i {\bf k} \cdot {\bf x}},\ \ \ \ \Psi({\bf x}, \tau)= \Psi_k(\tau)\  e^{i {\bf k} \cdot {\bf x}},
\ee
it is straightforward to get from Eq. \eqref{firstord}
\begin{align} \label{firstord1}
&\delta \phi_k^{\prime \prime}+ 2 \mathcal{H}\  \delta \phi_k^\prime + k^2 \delta \phi_k -4 \Psi_k^\prime \phi_k^\prime \notag \\
&= - \xi \left(2 k^2 \Psi_k-6\Psi^{\prime \prime}_k-24 \mathcal{H} \Psi^\prime_k-12 \frac{a^{\prime \prime}}{a} \Psi_k -4 k^2 \Psi_k  \right) \phi \notag \\
&\ \ \ -\left( V^{\rm eff}_{, \phi \phi}\ \delta \phi_k + 2 \Psi_k V^{\rm eff}_{, \phi } \right) a^2,
\end{align}
that turns out to be a complicated version of the equations of motion for $\phi$ at a perturbative level. Once the $\delta \phi_k$ modes are obtained, the full expansion for quantum fluctuations of the inflaton field reads \cite{ref1}
\begin{align} \label{fullexp}
\delta \hat{\phi} ({\bf x}, \tau)= \frac{1}{(2\pi)^{3/2}} \int d^3k \big( &\hat{a}_k \delta \phi_k(\tau) e^{i {\bf k} \cdot {\bf x}} \notag \\
&+ \hat{a}^\dagger_k \delta \phi_k^*(\tau) e^{-i {\bf k} \cdot {\bf x}}   \big),
\end{align}
where the ladder operators satisfy canonical commutation relations
\be \label{commrel}
[\hat{a}_k, \hat{a}_{k^\prime}^\dagger]= \delta^{(3)}({\bf k}-{\bf k^\prime}).
\ee
We will discuss normalization of the inflaton modes later on.
In order to solve analytically Eq. \eqref{firstord1}, one has to argue particular energy conditions, corresponding to given lengthscales for the inflaton fluctuations.

\subsection{Super-Hubble scales}

To leading order, each Fourier mode in Eq. \eqref{fullexp} evolves independently. The comoving Hubble radius during inflation,
\be \label{hrad}
r_H(\tau)=\frac{1}{a(\tau)H_I},
\ee
plays a key role in determining the mode dynamics: on \emph{sub-Hubble scales} ($k \geq a H_I$) the inflaton fluctuations typically oscillates, while they are nearly time-independent on \emph{super-Hubble scales} ($k < a H_I$), as we will see.

Formally, one can define an Hilbert space associated to fluctuation modes, which naturally divides into a sub-Hubble subspace and a super-Hubble one \cite{entaper}. Note that the comoving Hubble radius decreases as function of time during the inflationary period: this means that the boundary between the two subspaces depend on time, i.e., the more inflation goes on the more modes get off the horizon. This is a specific feature of systems on a dynamically expanding background.

It has been shown \cite{sq1,sq2,sq3} that some mixing may arise between sub- and super-Hubble modes, leading to decoherence of the reduced density matrix for both subsystems. However, these effects typically derive from interaction terms which are cubic in the perturbation variables \cite{entaper}, i.e., of higher order with respect to the field-geometry coupling studied here. For this reason, decoherence effects will be neglected in this work but will be subject of future investigations.

 In the context of GPP, it can be proven \cite{lyth} that particle production is dominant on super-Hubble scales, with respect to sub-Hubble ones, if one assumes a pure de Sitter evolution during inflation.  Accordingly, it seems interesting to generalize such a framework by including perturbations.

More precisely, the modes of interest are well inside the horizon at the beginning of inflation, and leave it, becoming super-Hubble, subsequently. This mechanism may also affect geometric particle production, as we will see. Moreover, the choice of such scales will naturally provide an infrared and ultraviolet cutoff for the momenta of the particles that will be produced.

The term $\Psi_k^\prime \phi_k^\prime$ in Eq. \eqref{firstord1} can be neglected on super-Hubble scales because perturbations are nearly frozen\footnote{Accordingly, a term of the form $\Psi_k^\prime \phi_k^\prime$ would be of higher order with respect to $\Psi_k\  V^{\rm eff}_{, \phi}$, since also $\Psi_k$ is small \cite{ref4} on these scales. The same reasoning apply to the terms $\Psi_k^\prime$ and $\Psi_k^{\prime \prime}$ in the curvature contribution.} in this limit. In this limit we also have
\be \label{pertein}
\Psi_k \simeq \epsilon \mathcal{H} \frac{\delta \phi_k}{\phi^\prime},
\ee
where
\be \label{slowroll}
\epsilon\equiv 1- \frac{\mathcal{H}^\prime}{\mathcal{H}^2} = 4 \pi G \frac{\phi^{\prime2}}{\mathcal{H}^2}
\ee
is the slow-roll parameter and $G$  the gravitational constant. Eq. \eqref{pertein} can be derived from the (0,i)-component of perturbed Einstein's equations \cite{ref4}.

Bearing the above considerations in mind, we can rewrite Eq. \eqref{firstord1} as
\begin{align} \label{eqpert2}
&\delta \phi_k^{\prime \prime}+ 2 \mathcal{H}\  \delta \phi_k^\prime+ \left[k^2+ \left( V^{\rm eff}_{, \phi \phi} + 2\epsilon \frac{\mathcal{H}}{\phi^\prime}\  V^{\rm eff}_{, \phi }\right) a^2  \right] \delta \phi_k \notag \\
&+\xi\left( -2k^2-12 \frac{a^{\prime \prime}}{a} \right) \Psi_k \phi =0.
\end{align}
For $\lvert \xi \rvert \ll 1$, we can neglect the contribution arising from the variation of the scalar curvature, since we also need the perturbation potential to satisfy $\lvert \Psi_k \rvert \ll 1$.

In the case of \emph{slow-roll approximation}, we can also set $\phi^{\prime \prime} \simeq 0$ and thus write the derivative of the potential as function of $\phi^\prime$ in the background equation, Eq. \eqref{zeroord}.

Performing now the usual rescaling procedure over the field \cite{ref3,ref4},
\be \label{sub}
\delta \phi_k \rightarrow \delta \chi_k = \delta \phi_k a,
\ee
and choosing the chaotic potential\footnote{Chaotic potentials  usually exhibit the \emph{graceful exits}, byproduct of attractor solutions as $\phi\rightarrow0$. The standard forms, namely $\sim \phi^2$ and $\sim \phi^4$, have been recently ruled out by the Planck satellite results \cite{planck} that, conversely, showed that they can work only if the curvature is coupled to $\phi$. We here limit ourselves to $\sim\phi^2$ in order to compute an analytic toy-model approach for entanglement production during inflation. More complicated cases invoke alternative potentials \cite{strutture} and will be object of future efforts.}
\be \label{chaotic}
V(\phi)= \frac{1}{2} m^2 \phi^2,
\ee
where $m$ is the mass of the field, we arrive at
\be \label{eqpert3}
\delta \chi_k^{\prime \prime}+ \left[k^2+m^2a^2- \left( 1-6\xi \right) \frac{a^{\prime \prime}}{a} - 6 \epsilon \left( \frac{a^\prime}{a} \right)^2  \right] \delta \chi_k=0.
\ee

%%%%%%%%%%%%%%%%%%%%%%%%%%%%%%%%%%%%%%%%%%%%%%%%%%%%%%%%%%%%%%%%%%%%%%%%
\section{Inflation in a \emph{quasi-de Sitter} spacetime} \label{sez3}

During inflation, the slow-roll parameters are small and almost constant \cite{iv1}. Commonly, one refers to this assuming that a suitable solution for the scale factor turns out to be purely de Sitter. However, this happens only in the simplest cases, i.e., when vacuum energy dominates \cite{iv2}. In fact, since vacuum energy is constant, the scale factor naturally reads as an exponential, implying a de Sitter phase. Clearly, for a generic potential, that does not reduce to vacuum energy during inflation, the situation is different. Indeed, one has to solve the equations of motion for the field and, by virtue of the Friedmann equations, arguing the exact form of $a(\tau)$ throughout inflation. This is clearly not easy and quite often appears as a sole numerical investigation.

Thus, during the inflationary stage one can approximate the scale factor through a \emph{quasi-de Sitter} function that appears to suitably approximate the real dynamics and the slow-roll parameter as well. In particular, following Ref. \cite{ref4}, we here propose the approximate quasi-de Sitter solution provided by
\be \label{scale}
a(\tau)= -\frac{1}{H_I} \frac{1}{\tau^{(1+v)}},\ \ \ \ \ v \ll 1,
\ee
where $\tau <0$ and $H_I$ is the Hubble parameter during inflation, up to small corrections. In this respect, Planck data impose severe upper bounds on $H_I$, leading to  \cite{planck}
\be
H_I/M_{\rm pl} \lesssim 2.5 \times 10^{-5},
\ee
where $M_{\rm pl}$ is the Planck mass. The parameter $v$ in Eq. \eqref{scale} essentially describes small deviations from a pure de Sitter phase. We notice that
\be \label{slow}
\epsilon= 1- \frac{\mathcal{H}^\prime}{\mathcal{H}^2}= 1- \frac{1}{1+v} \simeq v,
\ee
implying that we can identify $v$ as a small and constant slow-roll parameter. Departures from this approximate version of the scale factor with respect to the real solution for $a(\tau)$ are extremely small, in the slow-roll regime. Accordingly, we set $v \equiv \epsilon$ from now on.
Using now Eq. \eqref{scale} and noting that $a^{\prime \prime}/a \simeq (2+3\epsilon)/\tau^2$, the equation of motion for perturbations \eqref{eqpert3} finally gives
\be \label{eqpert4}
\delta \chi_k^{\prime \prime}+ \left[k^2-\frac{1}{\tau^2}\left( (1-6\xi)(2+3\epsilon)+6 \epsilon -\frac{m^2}{H_I^2} \right)\right] \delta \chi_k=0.
\ee
This equation can be recast in the form
\be \label{bess}
\delta \chi_k^{\prime \prime} + \left[ k^2- \frac{1}{\tau^2} \left( \nu^2-\frac{1}{4} \right)   \right] \delta \chi_k=0,
\ee
where
\be \label{hankind}
\nu^2= \frac{1}{4}+ (1-6\xi)(2+3\epsilon)+6\epsilon- \frac{m^2}{H_I^2}.
\ee
This result coincides with that of Ref. \cite{ref4} in the case of minimal coupling, $\xi=0$, if one introduces the further parameter $\eta=m^2/3 H_I^2$. For small $\xi$, i.e., $\xi \simeq (\epsilon; \eta)$,  it is easy to see that\footnote{More precisely, since the inflaton field is massive,  the condition $\lvert \xi \rvert \lesssim 10^{-3}$ is required during inflation, see e.g. \cite{fut, tsu2}.}
\be \label{index}
\nu \simeq \sqrt{\frac{9}{4}+9 \epsilon-3\eta-12 \xi} \simeq \frac{3}{2}+3\epsilon-\eta-4\xi.
\ee
The general solution of Eq. \eqref{bess} can be written in the form
\be \label{solbess}
\delta \chi_k(\tau)= \sqrt{-\tau} \left[ c_1(k) H_\nu^{(1)}(-k \tau)+ c_2(k) H_\nu^{(2)}(-k \tau) \right],
\ee
where $H_\nu^{(1)}$ and $H_\nu^{(2)}$ are Hankel functions and the constants $c_1(k)$ and $c_2(k)$ can be determined by imposing the normalized initial vacuum state.

A common choice is the \emph{Bunch-Davies vacuum} \cite{ref7,ref8,ref9}, i.e., to impose that in the ultraviolet regime $k \gg aH_I$ the solution for $\delta \chi_k$ matches the following plane-wave solution:
\begin{equation}
\delta \chi_k\sim e^{-ik\tau}/\sqrt{2k}\,.
\end{equation}
Thus, knowing that
\begin{subequations}
\begin{align}
& H_\nu^{(1)}(x \gg 1) \simeq \sqrt{\frac{2}{\pi x}} e^{i(x-\frac{\pi}{2}\nu-\frac{\pi}{4})},  \label{hankel1}\\
&H_\nu^{(2)}(x \gg 1) \simeq \sqrt{\frac{2}{\pi x}} e^{-i(x-\frac{\pi}{2}\nu-\frac{\pi}{4})}, \label{hankel2}
\end{align}
\end{subequations}
we can then set
\be \label{coeff}
c_1(k)=\frac{\sqrt{\pi}}{2}e^{i(\nu +\frac{1}{2})\frac{\pi}{2}},\ \ \ c_2(k)=0.
\ee
This gives the solution
\be \label{sol}
\delta \chi_k(\tau)= \frac{\sqrt{\pi}}{2}e^{i(\nu +\frac{1}{2})\frac{\pi}{2}} \sqrt{-\tau} H_\nu^{(1)}(-k \tau).
\ee
On super-Hubble scales, since $H_\nu^{(1)}(x \ll 1) \simeq \sqrt{2/\pi} e^{-i \pi/2}\  2^{\nu-3/2} \left( \Gamma(\nu)/\Gamma(3/2) \right) x^{-\nu}$, the fluctuation becomes
\be \label{solsh}
\delta \chi_k= e^{i \left( \nu- \frac{1}{2} \right) \frac{\pi}{2} } 2^{\left( \nu- \frac{3}{2} \right)} \frac{\Gamma(\nu)}{\Gamma(3/2)} \frac{1}{\sqrt{2k}} (-k \tau)^{\frac{1}{2}-\nu}.
\ee
\begin{figure} [ht!]
    \centering
    \includegraphics[scale=0.6]{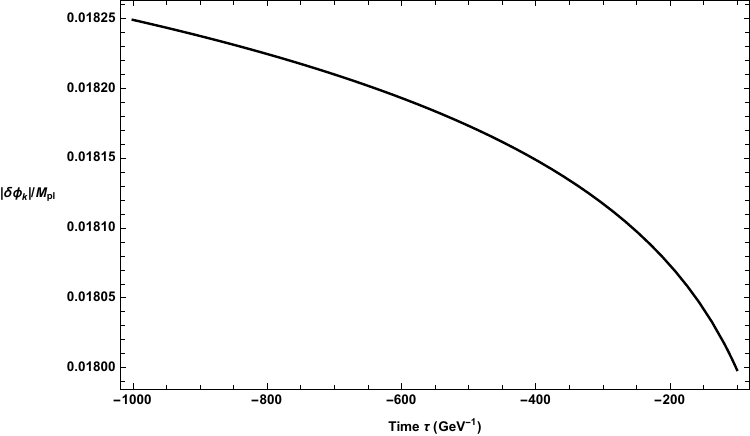}
        \caption{Fluctuations of the inflaton field $\lvert \delta \phi_k (\tau) \rvert$, normalized with respect to the Planck mass. The other parameters are: $H_I=10^{13}$ GeV, $\epsilon=10^{-3}$, $\xi=10^{-3}$ and $k\equiv0.001$, which corresponds to scales crossing the Hubble horizon at the beginning of inflation, in our model. We also set $\eta=5 \times 10^{-3}$, corresponding to an inflaton mass $m \simeq 1.22 \times 10^{12}$ GeV.}
    \label{fig1}
\end{figure}

Restoring now the original fluctuation $\delta \phi_k$, we obtain
\be \label{fluct}
\delta \phi_k=  e^{i\left(\nu-\frac{1}{2}\right)\frac{\pi}{2}}  2^{\nu-\frac{3}{2}} \frac{\Gamma(\nu)}{\Gamma(3/2)} \frac{H_I}{\sqrt{2k^3}} \left(\frac{k}{aH_I} \right)^{\frac{3}{2}-\nu},
\ee
which is plotted in Fig. \ref{fig1} as function of conformal time.

We remark that this result is correct only in the slow-roll regime, where the Universe expansion can be described by a scale factor of the form \eqref{scale}. The corresponding perturbation can be now derived from Eq.  \eqref{pertein}, once we solve Eq. \eqref{zeroord} for the background field.

Hence, including the slow-roll hypothesis, this equation gives
\begin{align} \label{backg}
2 \left( \frac{1+\epsilon}{\tau}  \right) \phi^\prime = a^2 \left(m^2+6\xi \frac{a^{\prime \prime}}{a^3} \right) \phi \simeq\\
\frac{1}{\tau^2} \left( 3 \eta+ 6 \xi (2+3\epsilon) \right) \phi,\notag
\end{align}
with solution
\begin{align} \label{backsol}
\phi(\tau)= c_0 \lvert \tau \rvert^{ \big(3 \eta+ 6 \xi (2+3\epsilon) \big)/\left( 2+2\epsilon \right)}=
c_0 \lvert \tau \rvert^\kappa,
\end{align}
where compactly
\begin{equation}
    \kappa \equiv \frac{3[ \eta+ 2 \xi (2+3\epsilon)]}{ 2(1+\epsilon)}\,.
\end{equation}
The integration constant $c_0$ can be determined by imposing the initial value of the background field $\phi(\tau_i)$, while the coupling constant $\xi$ is small, as previously discussed.

For $\xi \simeq (\epsilon; \eta)$ we can neglect second order terms and thus write
\be \label{simpl}
\kappa \simeq \frac{ 3( \eta + 4 \xi)}{2(1+\epsilon)}.
\ee
The initial and final times $\tau_i, \tau_f$ for the inflationary era can be derived by selecting a given number of e-foldings $N$. Commonly one takes  $N\gtrsim60$, i.e., those minimally needed to speed the Universe up during inflation,
\be \label{efoldings}
N= \int\ dt H(t) \simeq - \int_{\tau_i}^{\tau_f}\ d\tau\ \frac{H_I}{H_I \tau} = 60.
\ee
\begin{figure} [ht!]
    \centering
    \includegraphics[scale=0.6]{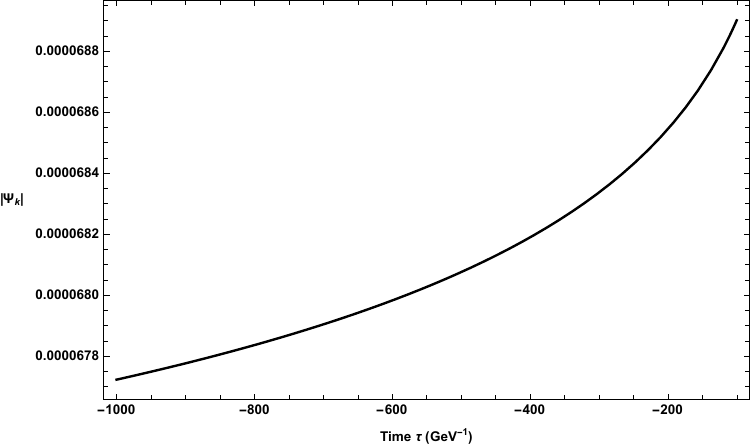}
    \caption{Perturbation potential $\lvert \Psi_{k_i}(\tau) \rvert$. The parameters chosen are:  $H_I=10^{13}$ GeV, $\epsilon=10^{-3}$, $\xi=10^{-3}$, $\eta=5 \times 10^{-3}$ and $\phi(\tau_i)=20$ M$_{\rm pl}$. }
    \label{fig2}
\end{figure}

Since we are focusing on modes exceeding the Hubble horizon after the beginning of inflation, we set $k>k_i= H_I a(\tau_i)$ and we further require the perturbation potential to be small with respect to the background, namely $ \lvert \Psi_k \rvert \ll 1\,,  \forall k$.

For instance, a viable choice is $\tau_i=-10^{3}$, that in turn gives $k_i=0.001$. Accordingly, via Eq. \eqref{efoldings} we can derive the corresponding value for $\tau_f$ and recalling the relation between the inflaton field value and the number of e-foldings\footnote{This equation is valid in case of chaotic potential, see \cite{ref5}. Our potential can be seen as chaotic, since the curvature is almost constant in the quasi-de Sitter phase, resulting in an effective mass $m^{\rm eff}= \sqrt{m^2+ \xi R}$, see Eqs. \eqref{effectivepot} and \eqref{chaotic}. Of course, this assumption is no longer valid as inflation ends, when the quasi de Sitter scale factor do not describe properly the Universe evolution.} before the end of inflation
\be
N(\phi) \simeq \frac{\phi^2}{4 M_{\rm pl}^2}-\frac{1}{2},
\ee
we can fix $\phi(\tau_i)$ and the corresponding value for $c_0$. The geometric perturbation, Eq. \eqref{pertein}, on super-Hubble scales finally takes the form
\begin{align} \label{geo}
\Psi_k(\tau)=& -\epsilon\ e^{i\left(\nu-\frac{1}{2}\right)\frac{\pi}{2}}  2^{\nu-\frac{3}{2}} \frac{\Gamma(\nu)}{\Gamma(3/2)}\ \frac{\mathcal{H}} {\big(\kappa c_0 \lvert \tau \rvert^{\kappa-1} \big)} \notag \\
& \times \frac{H_I}{\sqrt{2k^3}} \left(\frac{k}{aH_I} \right)^{\frac{3}{2}-\nu},
\end{align}
and it is plotted in Fig. \ref{fig2} as function of time, assuming $k=k_i$.

%%%%%%%%%%%%%%%%%%%%%%%%%%%%%%%%%%%%%%%%%%%%%%%%%%%%%%%%%%%%%%%%%%%%%%%%

\section{Particle production in inflationary phase} \label{sez4}

Once the geometric perturbation has been computed, we can quantify the amount of particles arising from the coupling of inflaton fluctuations $\delta \phi({\bf x}, \tau)$ to gravity. According to previous findings \cite{bel2},  we will call \emph{geometric particles} those quasi-particles obtained when the inflaton is coupled to the scalar curvature $R$.

Writing the perturbed metric in the form $g_{\mu \nu}=a^2(\tau)(\eta_{\mu \nu}+h_{\mu \nu})$, we can describe at first order the interaction between fluctuations and spacetime perturbations via the Lagrangian\footnote{ An alternative approach consists in describing both inflaton fluctuations $\delta \phi_k$ and scalar metric perturbations $\Psi_k$ via the Mukhanov-Sasaki variable \cite{muksak}. This choice turns out to be useful in some particular contexts, e.g. particle production during preheating \cite{georeh}.} \cite{ref1}
\be \label{lagint}
\mathcal{L}_I= -\frac{1}{2} \sqrt{-g_{(0)}} H^{\mu \nu}T_{\mu \nu}^{(0)}\,,
\ee
where $g_{\mu \nu}^{(0)}\equiv a^2(\tau) \eta_{\mu \nu}$, $H_{\mu \nu}=a^2(\tau)h_{\mu \nu}$ and $T_{\mu \nu}^{(0)}$ is the zero-order energy-momentum tensor associated to fluctuations, which is given by

\begin{align} \label{zerotens}
T_{\mu \nu}^{(0)}=& \partial_{\mu} \delta \phi\  \partial_{\nu}\delta \phi-\frac{1}{2} g_{\mu \nu}^{(0)} \left[ g^{\rho \sigma}_{(0)}\  \partial_{\rho} \delta \phi\  \partial_{\sigma} \delta \phi - m^2(\delta \phi)^2  \right] \notag \\
&- \xi \left[ \nabla_{\mu} \partial_{\nu}- g_{\mu \nu}^{(0)} \nabla^\rho \nabla_\rho+R_{\mu \nu}^{(0)}-\frac{1}{2} R^{(0)} g_{\mu \nu}^{(0)}   \right] (\delta \phi)^2.
\end{align}

Since the energy-momentum tensor is quadratic in the fluctuations, particles are produced in pairs at first perturbative order.

The corresponding number density of geometric particles produced at a given time $\tau^*$ is given by
\begin{align} \label{numbdens}
N^{(2)}(\tau^*)= \frac{a^{-3}(\tau^*)}{\left(2\pi  \right)^{3}} \int d^3q\  d^3p\  &\lvert \langle 0 \lvert \hat S \rvert p,q \rangle \rvert^2 \notag \\ &\times \left( 1+ \lvert \beta_p \rvert^2+ \lvert \beta_q \rvert^2 \right).
\end{align}
where $\beta_p$ and $\beta_q$ are Bogolubov coefficients \cite{gp1,ref1}, associated to the field evolution with respect to the homogeneous background. See, for example, \cite{bel2} for a derivation of Eq. \eqref{numbdens}. As discussed in the Introduction, nonzero Bogolubov coefficients leads to gravitational (or quantum) particle production (GPP), provided by a consolidate mechanism, see e.g. \cite{gp1,gp2,gp3, gp4}, and also widely-investigated in the inflationary regime \cite{lyth,gp5}. In the context of cosmological perturbations, the GPP process due to inflationary expansion is also associated to entropy generation, as the result of squeezing of fluctuations modes on super-Hubble scales \cite{entrop1,entrop2,entrop3}. As we will discuss later on, this mechanism differs from entanglement production due to perturbations only.

The main disadvantage in dealing with Bogolubov transformations on a FRW background is that they only mix modes of the same momentum \cite{ent2}. This leads, in principle, to particle-antiparticle pair production, which may annihilate. On the other side, geometric particle production is not restricted to such pairs. This is due to the presence of inhomogeneities, which break space translation symmetry so that linear momentum is no longer conserved.  In Eq. \eqref{numbdens} we notice the presence of a purely geometric contribution, namely the first term, but we also notice that nonzero Bogolubov coefficients can enhance the geometric mechanism of production here studied, resulting in a larger number of particles produced. This effect should be further investigated, especially in the attempt of deducing dark matter from a geometric mechanism of particle production \cite{bel2}. Alternative proposals for geometric quasi-particles have also been studied, see e.g. \cite{add1}.

Before discussing particle production associated to the inflaton fluctuations, we underline that the number density in Eq. \eqref{numbdens} can be computed analytically only in some special cases. For example, assuming a conformally coupled massless scalar field ($m=0$, $\xi=1/6$) it can be shown that the Bogoliubov coefficients are zero, i.e., the background expansion does not produce particles and the second order number density reduces to
\be \label{massless}
N^{(2)}= \frac{1}{960 \pi} \int d^4x\  a^4(\tau) C_{\mu \nu \rho \sigma} C^{\mu \nu \rho \sigma},
\ee
where $C_{\mu \nu \rho \sigma}$ is the Weyl tensor associated to the perturbed metric $g_{\mu \nu}$. Other examples are discussed in \cite{ref1}.

The $S$ matrix $\hat{S}$ in Eq. \eqref{numbdens} is derived from the first-order Dyson's expansion, namely
\be \label{dyson}
\hat S \simeq 1+i \hat{T} \int d^4x\ \mathcal{L}_I.
\ee
We remark that a proper definition of the $S$ matrix in curved spacetime is not straightforward \cite{gp3, wald, audr}. First of all, we need the interaction to be switched off in the distant past and future, as for Minkowski spacetime. In our model this assumption seems realistic, since in inflationary cosmology all pre-existing classical fluctuations can be typically neglected (see e.g. \cite{brandbook}), while at the end of the slow-roll regime we expect back-reaction effects to gradually restore homogeneity, as discussed in Sec. \ref{sez5}.

At the same time, we are faced with the problem of properly defining particle states, which is a peculiar issue of quantum field theory in curved spacetime. In a de Sitter background, which clearly does not possess asymptotically flat regions, a valid definition of initial no-particle states can be given in terms of the adiabatic vacuum \cite{gp3}. In particular, it can be shown that the Bunch-Davies vacuum introduced in Sec. \ref{sez3} is a local attractor in the space of initial states for an expanding background \cite{attract}.

In eternal de Sitter space one can prove that no particle production arises due to the background. However, the Universe dynamics is clearly not described by a scale factor of the form \eqref{scale} at any time and, more subtly, a de Sitter background still produces thermal radiation, which can be detected by comoving observers in it \cite{gp3, bousso}. For these reasons, a realistic description of spacetime evolution necessarily requires the inclusion of Bogolubov coefficients, as result of the transition from inflation to radiation/matter domination and then late times \cite{lyth}. In turn, this also implies an increase of the total amount of geometric particles produced, as shown by Eq. \eqref{numbdens}. We will deepen this point in future works.

\begin{figure}[ht!]
  \includegraphics[scale=0.6]{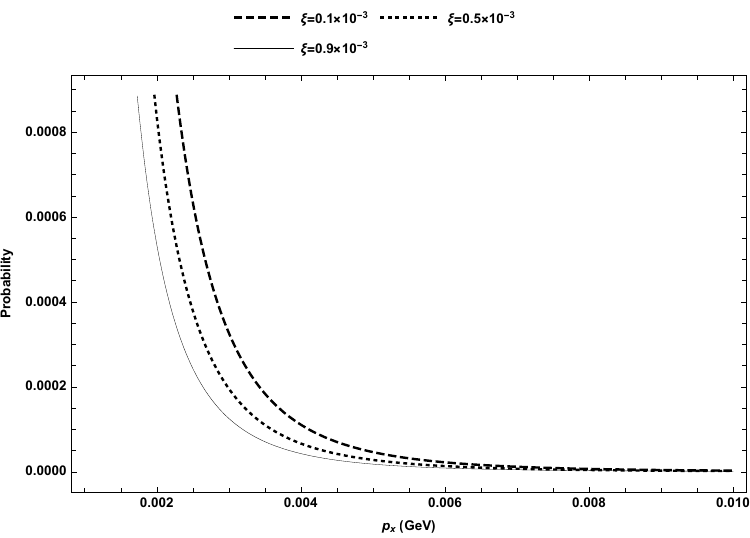}
  \includegraphics[scale=0.6]{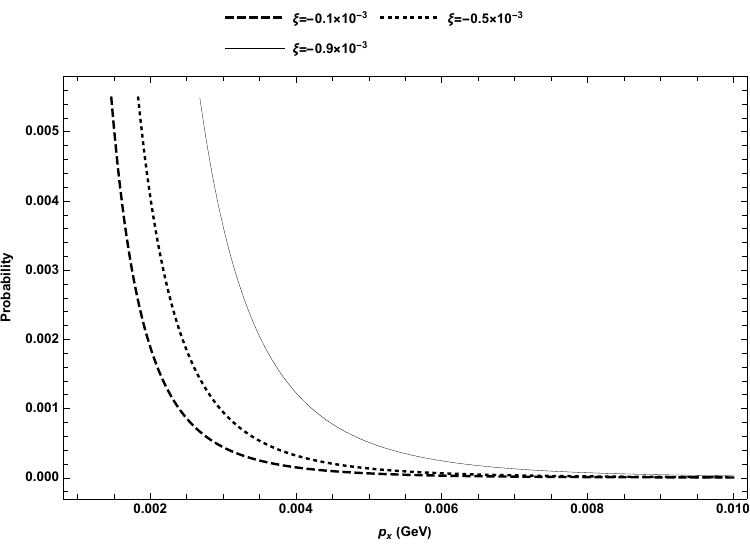}
  \caption{Pair production probability $\lvert \langle p,q \rvert \hat{S} \rvert 0 \rangle \rvert^2$ as function of the momentum $p_x$, for positive and negative values of the coupling parameter $\xi$. We have assumed $q_x=0.01$ GeV and $p_y=p_z=0$, while the other parameters are the same as in Fig. \ref{fig2}.}
  \label{fig3}
\end{figure}

For the moment, we focus on the probability amplitude
\begin{widetext}
\begin{align} \label{probam}
\langle p,q \lvert \hat S \rvert 0 \rangle=& -\frac{i}{2 \mathcal{N}} \int d^4x\ 2a^4  H^{\mu \nu} \bigg[ \partial_{(\mu} \delta \phi_p^* \partial_{\nu)}\delta \phi_q^*-\frac{1}{2}\eta_{\mu \nu} \eta^{\rho \sigma} \partial_{(\rho} \delta \phi_p^* \partial_{\sigma)} \delta \phi_q^*+\frac{1}{2}g_{\mu \nu}^{(0)} m^2 \delta \phi_p^* \delta \phi_q^*  \notag \\
&\ \ \ \ \ \ \ \ \ \ \ \ \ \ \ \ \ \ \ \ \ \ \ \ -\xi \left( \nabla_\mu \partial_\nu-g_{\mu \nu}^{(0)}\nabla^\rho \nabla_\rho+R_{\mu \nu}^{(0)}-\frac{1}{2} R^{(0)}g_{\mu \nu}^{(0)} \right)\delta \phi_p^* \delta \phi_q^* \bigg] e^{-i({\bf p}+{\bf q})\cdot {\bf x}}\,,
\end{align}
\end{widetext}
where $\mathcal{N}$ is a normalization factor. Exploiting the fact that the perturbation tensor is diagonal and writing explicitly all the curvature terms, Eq. \eqref{probam} can be expressed as
\begin{align} \label{compact}
\langle p,q \lvert \hat S \rvert 0 \rangle= -\frac{i}{2 \mathcal{N}} \int d^4x\  2a^2\big(&A_0({\bf x}, \tau)
+A_1({\bf x},\tau) \notag \\
&+A_2({\bf x},\tau)+A_3({\bf x},\tau) \big),
\end{align}
where
\begin{align} \label{A0}
A_0({\bf x},\tau)= 2 \Psi \bigg[&\partial_0 \delta \phi_p^*\  \partial_0 \delta \phi_q^*-\frac{1}{2} \eta^{\rho \sigma} \partial_\rho \delta \phi_p^* \ \partial_\sigma \delta \phi_q^* \notag \\
    &+\frac{1}{2} m^2 a^2 \delta \phi_p^* \delta \phi_q^*- \xi \bigg(\partial_0 \partial_0-\frac{a^\prime}{a}\partial_0 \notag \\
    &-\eta^{\rho \sigma}\partial_\rho \partial_\sigma-3\left( \frac{a^\prime}{a} \bigg)^2   \right) \delta \phi_p^* \delta\phi_q^* \bigg] e^{-i({\bf p}+{\bf q})\cdot {\bf x}}\notag\\
\end{align}
and
\begin{align} \label{eq65}
    A_i({\bf x},\tau)=2 \Psi \bigg[&\partial_i \delta \phi_p^*\  \partial_i \delta \phi_q^*+\frac{1}{2} \eta^{\rho \sigma} \partial_\rho \delta \phi_p^*\  \partial_\sigma \delta \phi_q^* \notag \\
    &-\frac{1}{2} m^2 a^2 \delta \phi_p^* \delta \phi_q^*- \xi \bigg(\partial_i \partial_i+\frac{3a^\prime}{a}\partial_0+ \frac{2a^{\prime \prime}}{a} \notag \\
    &+\eta^{\rho \sigma}\partial_\rho \partial_\sigma-\left( \frac{a^\prime}{a} \right)^2  \bigg) \delta \phi_p^* \delta\phi_q^* \bigg] e^{-i({\bf p}+{\bf q})\cdot {\bf x}}.\notag\\
\end{align}

for $i=1,2,3$. Recalling Eq. \eqref{ansatz} for the perturbation potential, the integral over space leads to a Dirac delta. Moreover, time integration has to be performed so that all the modes of interest are in super-Hubble form, Eq. \eqref{fluct}. In particular, we evaluated particle production in the time interval $\tau \in [\tau^*, \tau_f]$, with $\tau^*=\tau_i/1000$. Such a choice ensures that all modes in the range $0.001 \leq k < 1$ lie within super-Hubble scales during this interval.
In Fig. \ref{fig3} we show the probability of particle production as function of the momentum $p_x$, assuming $p_y=p_z=0$ and $q=q_x=0.01$ GeV.

\subsection{Geometric particle production and warm inflation}

Particle production during inflation, which we quantified in the context of cosmological perturbations, is also a peculiar property of the so-called \emph{warm inflation} scenario\footnote{Warm inflation represents an alternative to the more popular cold inflation scenario. It allows for interactions between the inflaton and other quantum fields within the slow-roll regime, which are not present in the standard picture of inflation (denoted as ``cold"  for this reason).  We will not discuss technical aspects or models of warm inflation here: the interested reader may consult seminal papers on this topic \cite{warm1,warm2,warm3,warm4,warm5}, while more recent developments are summarized in the review \cite{warm6} and references therein.}. In this framework, one typically assumes that the interaction between the inflaton and radiation fields leads to dissipative effects, which can be interpreted in terms of particle production \cite{warm7, warm8}. During slow-roll, the evolution of the background inflaton field in warm inflation is described (in cosmic time) by \cite{warm7}
\be \label{warm}
(3H+ \Gamma) \dot{\phi}+ V_{,\phi} \simeq 0,
\ee
where for simplicity we have assumed no field-curvature coupling. The coefficient $\Gamma$ quantifies the effects of dissipation due to interactions, i.e., the energy transferred from the inflaton to other fields. Such term can be derived assuming some specific interaction Lagrangians, by means of thermal quantum field theory \cite{warm6, warm8}. However, an intuitive estimate of the dissipation rate can be obtained following Ref. \cite{warm2}: for a given interaction, we can argue that the dissipation rate is proportional to the probability of pair production and the corresponding background temperature when such process occurs. From Fig. \ref{fig3}, we notice that, in case of geometric production, probability amplitudes are usually low, except for modes that exit the horizon at the beginning of inflation. Even lower amplitudes are expected if the zero-order energy-momentum tensor, Eq. \eqref{zerotens}, is not associated to the inflaton fluctuations (e.g. for radiation or other scalar/fermionic fields), as consequence of the fact that interactions are purely gravitational in our model. Similarly, we expect the production rate to be negligible for sub-Hubble modes, due to the oscillatory behaviour of the inflaton fluctuations in that regime.

This suggests that geometric particle production cannot account for large dissipation rates $\Gamma \gtrsim H$ during inflation, at least in a single field scenario. However, many successful warm inflation models propose a two-stage mechanism, where the inflaton interacts with heavy intermediate fields, that in turn are coupled to light fields (either fermions or bosons) \cite{warm9}. Possible extensions of the here-proposed model to multifield inflation could then shed further light on the dissipative effects associated to geometric particle production.

Finally, we remark that dissipation could be also interpreted in terms of \emph{back-reaction} of the produced particles on the background field dynamics\footnote{The topic of back-reaction in cosmological perturbation theory has been widely investigated in recent years, see e.g. \cite{reviewback} for a review of the different techniques adopted. However, it has been shown  that in some cases \cite{boyan} back-reaction from particle production cannot be described by an interaction term of the form $\Gamma \dot{\phi}$.}. In the next section we discuss, from a classical perspective, back-reaction due to the inflaton fluctuations and the associated metric perturbations.

%%%%%%%%%%%%%%%%%%%%%%%%%%%%%%%%%%%%%%%%%%%%%%%%%%%%%%%%%%%%%%%%%%%%%%%%%

\section{Back-reaction effects and consequences on the energy-momentum tensor} \label{sez5}

The particle production mechanism discussed in Sec. \ref{sez4} is based on the so-called \emph{external field approximation}, i.e,  once the geometric perturbation has been computed, we evaluate the corresponding probability of pair production in this fixed (perturbed) background. However, as already noted in \cite{ref1}, we expect metric perturbations to alter the background evolution of the Universe, in such a way to reduce the particle creation rate. Accordingly, such \emph{back-reaction} effects should be taken into account in order to properly deal with the dynamics of a perturbed spacetime.

Back-reaction associated to density inhomogeneities was first studied in \cite{fut1,fut2}, focusing on its effects on local observables, such as the expansion rate of the Universe. A further step was the formulation of the \emph{classical}\footnote{As it will be clear soon, in the following we will deal with both metric and matter perturbations at a classical level, i.e., introducing a generalized variable $\delta q$ to describe inhomogeneities. For a semiclassical treatment of back-reaction in a quasi de Sitter spacetime, see for example \cite{fin1,fin2}.} back-reaction problem in a gauge-invariant way \cite{abr1,abr2}: this can be done via the introduction of an effective energy-momentum tensor (EEMT) for cosmological perturbations.

Following the notation of \cite{abr1}, we start by denoting the metric, $g_{\mu \nu}$, and matter, $\phi$, fields by the \emph{collective variable} $q^a$. Accordingly, we can write
\be \label{varq}
q^a=q_0^q+ \delta q^a,
\ee
where the background field $q_0^a$ is defined as the homogeneous part of $q^a$ on the hypersurfaces of constant time, while the perturbations $\delta q^a$ depend both on time and spatial coordinates and satisfy $\lvert \delta q^a \rvert \ll q^a_0$.

From Figs. \ref{fig1}-\ref{fig2} we clearly see that this assumption is satisfied both for metric and matter perturbations in our case. We also require
\be \label{avg}
\langle \delta q^a  \rangle= \frac{\int_V \delta q^a\ d^3x}{\int_V d^3x} =0,
\ee
where the above describes a \emph{spatial averaging}, defined with respect to the background metric.

Denoting the Einstein equations by
\be \label{einstein}
G_{\mu \nu}-8\pi G T_{\mu \nu} \coloneqq \Pi_{\mu \nu},
\ee
we can expand the tensor $\Pi_{\mu \nu}$ in a functional power series \cite{abr1} in powers of $\delta q^a$ around the background $q_0^a$, if we treat $G_{\mu \nu}$ and $T_{\mu \nu}$ as functionals of $q^a$.

Thus, we have
\be \label{expans}
\Pi(q_0^a)+\Pi_{,a} \big \rvert_{q_0^a} \delta q^a+ \frac{1}{2} \Pi_{,ab} \big \rvert_{q_0^a} \delta q^a \delta q^b+ \mathcal O(\delta q^3)=0.
\ee
Taking the spatial average of Eq. \eqref{expans} we obtain the corrected equations, which take into account the back-reaction of small perturbations on the evolution of the background, say
\be \label{avgbac}
\Pi(q_0^a)= - \frac{1}{2} \langle \Pi_{,ab} \delta q^a \delta q^b   \rangle.
\ee
We can require the latter expression to be gauge-invariant. To so so, we can introduce the new variable

\begin{equation}
Q=e^{\mathcal{L}_X}q\,,
\end{equation}
where $\mathcal{L}_X$ denotes a Lie derivative and is $X$ is constructed as a linear combination of the perturbation variables in Eq.\eqref{scalpert}, as shown in \cite{abr1,abr2}.
Accordingly, we define
\be \label{effemt}
\tau_{\mu \nu}(\delta Q)= - \frac{1}{16 \pi G} \langle \Pi_{\mu \nu, ab} \delta Q^a \delta Q^b  \rangle\,,
\ee
which is the gauge-invariant EEMT for cosmological perturbations.

\subsection{Back-reaction in inflationary regimes}

In the inflationary Universe scenario, the EEMT separates into two independent pieces, the first due to scalar perturbations and the second due to tensor modes\footnote{As it is well-known, vector modes decay in an expanding Universe, so they can be neglected in our analysis.}
\be \label{scaltens}
\tau_{\mu \nu}(\delta Q)= \tau_{\mu \nu}^{\rm scalar}(\delta Q)+\tau^{\rm tensor}_{\mu \nu}(\delta Q).
\ee

We focus on the scalar contribution and exploit gauge invariance to move to the longitudinal gauge. As discussed in Sec. \ref{sez2}, for a scalar field the variable $\Psi$ entirely characterizes metric perturbations, in this gauge. Under the slow-roll assumption, when dealing with super-Hubble perturbations the following results are obtained, as function of cosmic time:
\begin{align} \label{eemt}
    &\tau_{00} \simeq \frac{1}{2} V^{\rm eff}_{, \phi \phi}\  \langle \delta \phi^2 \rangle + 2V^{\rm eff}_{, \phi }\  \langle \Psi\  \delta \phi  \rangle, \\
    & \tau_{ij} \simeq a^2 \delta_{ij} \left[ \frac{3}{\pi G} H^2(t) \langle \Psi^2 \rangle- \frac{1}{2} V^{\rm eff}_{, \phi \phi}\  \langle \delta \phi^2 \rangle+ 2 V^{\rm eff}_{, \phi }\  \langle \Psi \delta \psi \rangle  \right].
\end{align}
Moving to conformal time, the energy density associated to back-reaction is then
\be \label{denback}
\rho_{\rm br} \equiv \tau^0_0 \simeq \left( \frac{2V^{\rm eff}_{,\phi \phi}\ (V^{\rm eff})^2}{(V^{\rm eff}_{, \phi})^2}-4V^{\rm eff} \right) \langle \Psi^2(\tau) \rangle.
\ee
and similarly one finds for the pressure $p_{\rm br}=-1/3\  \tau^i_i \simeq -\rho_{br}$.

The correlator $\langle \Psi^2 \rangle$ is given by \cite{abr2}
\be \label{corr}
\langle \Psi^2(\tau) \rangle= \int_{k_i}^{k_f} \frac{dk}{k} \lvert \Psi_k \rvert^2,
\ee
where the modes have been computed in Eq. \eqref{geo}. The integral runs over all modes with scales larger than the Hubble radius, i.e.,
\be \label{kf}
k< k_f(\tau) = H_I a(\tau),
\ee
but smaller than the Hubble radius at initial time $\tau_i$,
\be
k> k_i = H_I a(\tau_i),
\ee
namely all the modes that exit the Hubble horizon after the beginning of inflation (super-Hubble scales).

The effects of back-reaction can be then quantified by considering the fractional contribution of (scalar) perturbations to the total energy density: $\rho_{\rm br}/\rho_0$, where $\rho_0 \simeq V^{\rm eff}$ is the background energy density of the scalar field $\phi$.

\begin{figure}[ht!]
       \centering
\includegraphics[scale=0.6]{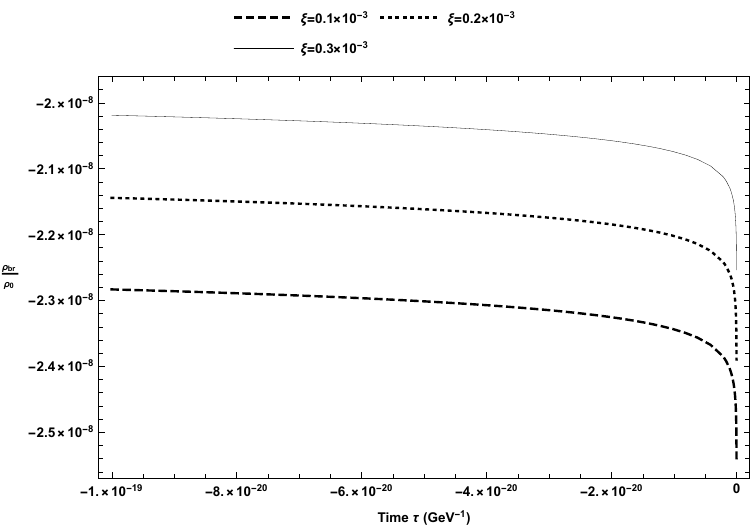}
\includegraphics[scale=0.6]{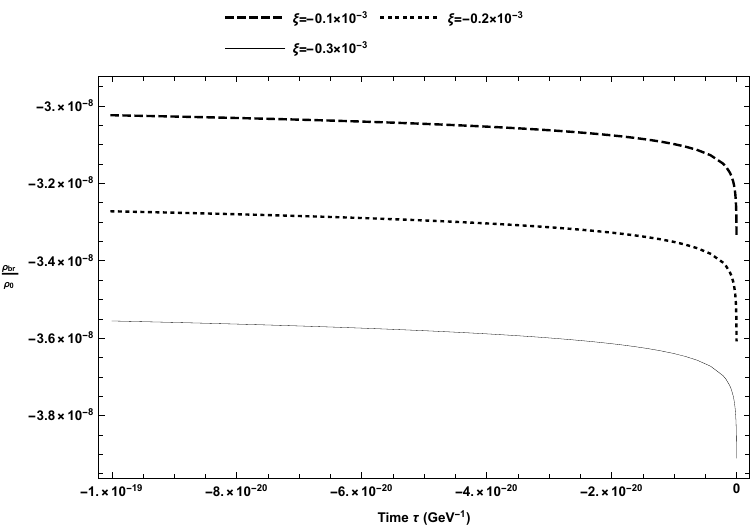}
\caption{Ratio $\rho_{\rm br}(\tau)/\rho_0$ for positive and negative values of the coupling parameter $\xi$, at times $\tau \lesssim \tau_f$, i.e., close to the end of the slow-roll regime. The other parameters are the same as in Fig. \ref{fig2}. The contribution of back-reaction is very small in both cases, so it can be neglected when dealing with geometric particle production.}
\label{fig4}
\end{figure}

In Fig. \ref{fig4}, the contribution of back-reaction is plotted for both positive and negative values of the coupling parameter $\xi$. As expected, $\rho_{\rm br} <0$, so \emph{back-reaction reduces the total amount of geometric particles produced}, since it gives a negative contribution to the zero-order energy-momentum tensor, Eq. \eqref{zerotens}.

However, its effects are almost negligible in the whole slow-roll phase, so it can be safely neglected when dealing with particle production during inflation. In other words, the effects of back-reaction does not significantly influence the net geometric particle production during the inflationary regime.

We also remark that back-reaction effects disappear in the limiting case of a pure de Sitter expansion, as due to $\epsilon=0$. This result appears evident, since for a pure de Sitter phase no particle production is possible at a perturbative level. The net effect would therefore be not to produce particles, but rather only to accelerate the Universe.

However, the above considerations do not enable one to ignore back-reaction at all stages of primordial Universe. Indeed, we expect back-reaction to play a more relevant role as the slow-roll approximation is no longer valid, i.e., close to the transition to reheating \cite{abr2}. In that epoch, therefore, baryon production appears to be dominant in fulfillment of the standard picture of reheating.

%%%%%%%%%%%%%%%%%%%%%%%%%%%%%%%%%%%%%%%%%%%%%%%%%%%%%%%%%%%%%%%%%%%%%%%%%

\section{Entanglement production at primordial time}  \label{sez6}

We finally quantify the entanglement entropy arising from geometric particle production at second order in the perturbation, i.e., when a purely geometric contribution is present. We follow the same approach introduced in Ref. \cite{bel1} and, as anticipated, we set here $\beta_p=\beta_q=0$. In this way we neglect the contribution due to GPP, that is typically interpreted in term of squeezing entropy between $k$ and $-k$ modes. This entropy has been widely investigated in cosmological scenarios, as discussed in the Introduction and more specifically in Sec. \ref{sez4} for the case of inflaton fluctuations. Crucially, it does not depend on the interaction, simply arising from the fact that the definition of positive and negative frequency modes typically differs between asymptotic in and out regions. Neglecting Bogolubov transformations, the $S$ matrix \eqref{dyson} gives the following final state of the system
\be \label{final}
\lvert \Phi \rangle = \hat{S} \lvert 0_p ; 0_q \rangle= \mathcal{N} \left(  \lvert 0_p ; 0_q \rangle+ \frac{1}{2} S^{(1)}_{pq} \lvert 1_p; 1_q \rangle  \right),
\ee
where we have introduced the shorthand notation $\langle p,q \lvert \hat{S} \rvert 0 \rangle \equiv S^{(1)}_{pq}$ and the constant $\mathcal{N}$ is derived from $\langle \Phi \rvert \Phi \rangle=1$.
\begin{figure}[ht!]
 \includegraphics[scale=0.6]{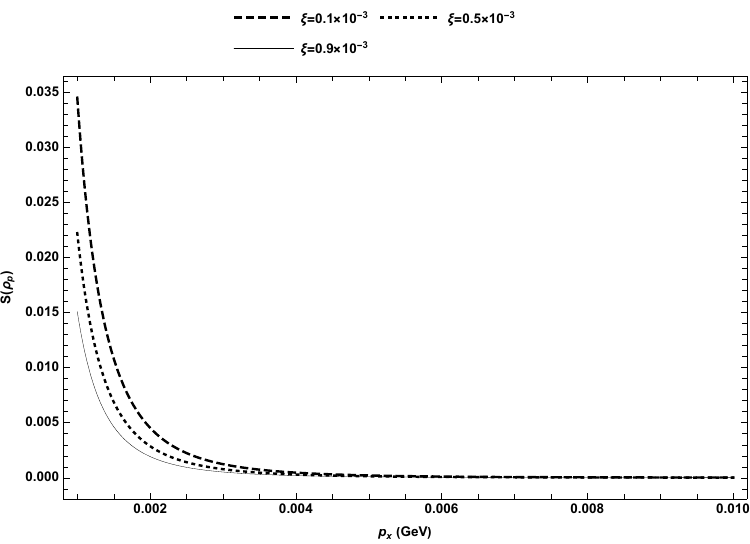}
\includegraphics[scale=0.6]{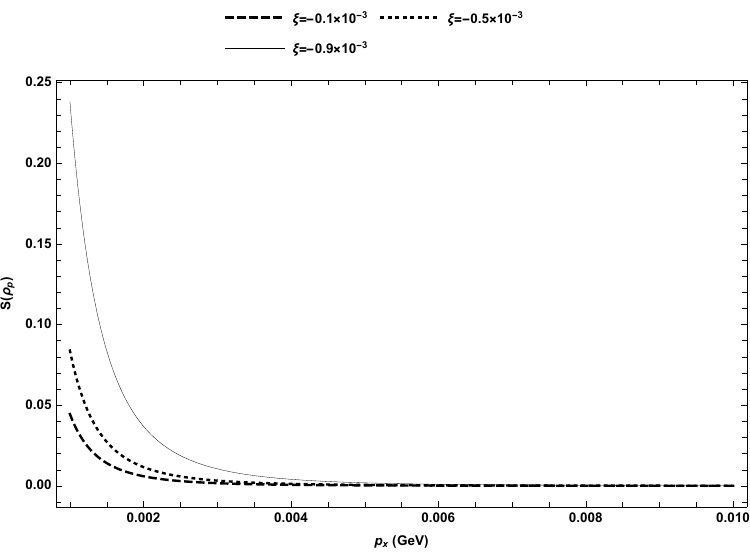}
\caption{Entanglement entropy of the reduced density operator $\rho_p$ as function of the momentum $p_x$, for positive and negative values of the coupling parameter $\xi$. The other parameters are the same as in Figs. \ref{fig2}-\ref{fig3}. Entanglement generation is higher in case of negative coupling constant, increasing at larger $\lvert \xi \rvert$.}
\label{fig5}
\end{figure}

Eq. \eqref{final} describes a bipartite pure state, whose entanglement entropy is quantified as usual by the von Neumann entropy of the reduced state, obtained after tracing out the $p$ or $q$ modes. Accordingly, the reduced density operator for the state \eqref{final} takes the form
\be \label{redop}
\rho_p = \text{Tr}_q (\lvert \Phi \rangle \langle \Phi \rvert)= \mathcal{N}^2 \left( \lvert 0 \rangle_p \langle 0 \rvert + \frac{1}{4} \lvert S^{(1)}_{pq} \rvert^2\  \lvert 1 \rangle_p \langle 1 \rvert   \right)
\ee
where the probability of pair production $\lvert S^{(1)}_{pq} \rvert^2$ is derived from Eqs. \eqref{compact} -- \eqref{eq65}, as discussed in Sec. \ref{sez4}.

The corresponding von Neumann entropy $\mathcal{S}(\rho_p)$ is plotted in Fig. \ref{fig5} as function of the momentum $p_x$, assuming again for simplicity that both particles are produced on the x-axis.

In analogy with the entanglement entropy associated to GPP \cite{ent2, ball,luo}, we notice that entanglement generation is higher as $p \rightarrow p_i$, where $p_i$ is defined as $k_i$ in Sec. \ref{sez3}. This is due to the bosonic nature of the field, for which modes of smaller $p$ are more easily excited, as expected.
The main difference, as discussed above, is that geometric particle production allows mode-mixing, thus leading to entanglement between particle pairs with $q \neq -p$.

Another crucial point is the following: for scalar fields, entanglement due to GPP  arises as consequence of the fact that the final state of the system is in the form of independent two-mode squeezed states \cite{ent2}. On the contrary, in our framework the evolution of the initial vacuum is governed by the $S$ matrix, that leads to the final state \eqref{final}. So, despite the mode dependence of the entanglement entropy is similar in both scenarios, the origin of such entropy turns out to be completely different. We remark that going beyond first order in Eq. \eqref{final} would imply that particle production is not limited to pairs, thus enriching the overall picture of \emph{geometric cosmological entanglement.} The same consideration applies if non-quadratic (e.g. quartic) potentials are chosen to describe the inflaton dynamics in place of Eq. \eqref{chaotic}, suggesting that the inflaton potential may significantly affect mode dependence of geometric entropy. Since our approach to inhomogeneities is a perturbative one, we notice that the amount of such geometric entanglement is typically small in our model: possible extensions to fully inhomogeneous spacetimes may shed further light on the properties of cosmological entanglement. We also notice that entanglement entropy is sensitive to the sign of the coupling constant between the field and the scalar curvature. This may be of crucial importance in understanding the nature of such coupling.

In fact, changing the coupling constant in the interacting potential can be interpreted as modifying the type of interaction between the scalar field and the gravity sector. Indeed, the $\xi$ positive sign corresponds to the \emph{attractive} behavior of the Yukawa-like contribution to the effective potential. Hence, shifting from positive to negative signs in the Yukawa contribution may lead to repulsive gravity effects and, in such a way, we can justify the deep difference that occurs as $\xi$ is modified. Repulsive gravity effects are not so rare in cosmological scenarios. For instance, dark energy models and/or extended theories of gravity seem to show similar effects \cite{repulsive1}. Analogously, in black holes and naked singularities often repulsive gravity are predicted to occur \cite{repulsive2,repulsive3}.

%%%%%%%%%%%%%%%%%%%%%%%%%%%%%%%%%%%%%%%%%%%%%%%%%%%%%%%%%%%%%%%%%%%%%%%%%

\section{Conclusions and perspectives}\label{conclusioni}

In this paper, we quantified the entanglement entropy associated to geometric particle production, specializing to the second order of perturbative expansion, i.e., assuming a purely geometric contribution. To do so, we adopted a single-field inflationary scenario, where the inflaton fluctuations are responsible for metric perturbations and also leads to back-reaction effects, studied from a classical point of view.

We investigated the dynamics of these fluctuations, understanding how they are responsible for the geometric mechanism of particle production, conjecturing these particles to contribute to dark matter abundance in the very early Universe.

We evaluated the modes and the corresponding analytical solutions for the inflaton field. The here-involved potential is a quadratic chaotic one, coupled to the scalar curvature. The corresponding effective potential is investigated and we computed the number of e-foldings, employing a quasi-de Sitter scale factor for the dynamics.

We studied then particle production and back-reaction effects. So, taking zero Bogolubov coefficients at first order expansion, we showed that the corresponding geometric particles and their probabilities for positive and negative coupling constants, $\xi$, are not deeply influenced by back-reaction effects. In fact, to show that, we got the amplitude element, adopting the Dyson expansion over the $S$ matrix, quantifying couples of particles with different momenta, in the limit of super-Hubble scales. We also compared geometric production rates to realistic dissipation rates in warm inflation scenarios, where the interaction of the inflaton with other quantum fields in  the slow-roll regime leads to particle production. We argued that tracing back such production to geometric effects would not lead to sufficient dissipation in a single field inflationary scenario.

Afterwards, we modeled the entropy of entanglement as due to the mode mixing of the above-obtained expansion. We showed its mode dependence and we focused on physical consequences on inflationary dynamics.

In analogy with the entanglement entropy associated to GPP \cite{ent2, ball,luo}, we noticed that entanglement generation is higher as $p \rightarrow p_i$, where $p_i$ is defined as $k_i$, as a consequence of the bosonic nature of the field itself, for which modes of smaller $p$ are more easily excited, as expected.

However, entanglement generated by the sole expansion of the Universe  has a different nature, because in this case the asymptotic out state of the system can be written as independent two-mode squeezed states, while inhomogeneities excite the initial vacuum only in terms of particle pairs. Consequently, we emphasized that the origin of such entropy turns out to be completely different.

The presence of inhomogeneities in the early Universe cannot be neglected, since these fluctuations are the seeds of cosmic structure. Accordingly, a complete characterization of cosmological entanglement cannot ignore spacetime perturbations. In particular, we demonstrated that the entropy due to geometric particle production is sensitive to the details of the expansion, e.g. to the initial value of the inflaton field and the Hubble parameter during inflation. This means that geometric cosmological entanglement may be useful in deducing some parameters which were crucial for the Universe evolution.

The latter is true in particular if the particle candidate in our model can be interpreted as dark matter, which is expected to have negligible interaction with standard matter: in this case residual quantum correlations may have survived to present time.

In general, our perturbative approach furnished a small correction under the form of geometric entanglement, as a consequence of how we treated inhomogeneities. Our model can be refined by including also the contribution due to Bogolubov coefficients at second perturbative order for particle production, which is expected to increase the total amount of entanglement.

Future works will also shed light on how to quantify entanglement in non-perturbative inhomogeneous contexts. Moreover, we will discuss additional  properties of cosmological entanglement, changing both the effective potential, likely considering more realistic ones, and the spacetime, assuming inhomogeneous solutions, instead of perturbing FRW. Finally, we will investigate more carefully the role played by such geometric production in dark matter scenarios, also including back-reaction effects both from a classical and semi-classical point of view.

\section*{Acknowledgments}
OL is grateful to the Department of Physics of the Al-Farabi University for hospitality during the period in which this manuscript has been written. This research has been partially funded by the Science Committee of the Ministry of Science and Higher Education of the Republic of Kazakhstan (Grant No. AP08052311).

\vskip 0.1in
%%%%%%%%%%%%%%%%%%%%%%%%%%%%%%%%%%%%%%%%%%%%%%%%%%%%%%%%%%%%%%%%%%%%%%%%%

\appendix

\section{Particle production in the synchronous gauge} \label{appA}
In this appendix we discuss geometric particle production in the synchronous gauge \cite{ref1,ref6}, where the most general scalar perturbation takes the form $h_{ij}^S=h \delta_{ij}/3+h_{ij}^\parallel$. The general procedure to transform from the longitudinal to the synchronous gauge is the following \cite{ref10}.
Let us consider a general coordinate transformation from a system $x^\mu$ to another $\hat{x}^\mu$
\be \label{S1}
x^\mu \rightarrow \hat{x}^\mu=x^\mu+d^\mu (x^\nu).
\ee
We write the time and the spatial parts separately as
\begin{subequations}
\begin{align}
&\hat{x}^0= x^0 + \alpha({\bf x},\tau) \label{S2} \\
&\hat{{\bf x}}= {\bf x}+ \nabla \beta({\bf x}, \tau)+ {\boldsymbol \epsilon}({\bf x}, \tau),\ \ \ \ \ \ \ \nabla \cdot {\boldsymbol \epsilon}=0, \label{S3}
\end{align}
\end{subequations}
where the vector $d$ has been divided into a longitudinal component $\nabla \beta$ and a transverse component $\vec{\epsilon}$.
Let $\hat{x}^\mu$ denote the synchronous coordinates and $x^\mu$ the conformal Newtonian coordinates, with $\hat{x}^\mu=x^\mu+d^\mu$. We have
\begin{subequations}
\begin{align}
& \alpha({\bf x}, \tau)=\beta^\prime ({\bf x}, \tau), \label{S4} \\
& \epsilon_i({\bf x}, \tau)=\epsilon_i({\bf x}), \label{S5} \\
& h_{ij}^\parallel({\bf x}, \tau)=-2 \bigg( \partial_i \partial_j-\frac{1}{3} \delta_{ij} \nabla^2 \bigg) \beta({\bf x}, \tau), \label{S6} \\
& \partial_i \epsilon_j+ \partial_j \epsilon_i=0. \label{S7}
\end{align}
\end{subequations}
and
\begin{subequations}
\begin{align}
    & \Psi({\bf x}, \tau)= -\beta^{\prime \prime} ({\bf x}, \tau)- \frac{a^\prime}{a} \beta^\prime ({\bf x}, \tau), \label{S8}\\
    & \Phi({\bf x}, \tau)= +\frac{1}{6} h({\bf x}, \tau)+ \frac{1}{3} \nabla^2 \beta({\bf x}, \tau)+ \frac{a^\prime}{a} \beta^\prime({\bf x}, \tau) \label{S9},
\end{align}
\end{subequations}
where $\Phi$ and $\Psi$ are the perturbation potentials in the longitudinal gauge. Now, Eq. \eqref{S8} gives
\be \label{betapert}
\beta^{\prime \prime}- \frac{1+\epsilon}{\tau} \beta^{\prime}+ \Psi_k e^{i {\bf k} \cdot {\bf x}}=0.
\ee
From Eqs. \eqref{ansatz} and \eqref{geo} we see that the geometric perturbation $\Psi_k$ is polynomial in time, i.e., it can be written as
\be \label{perprop}
\Psi = A_k  \left(-\tau\right)^{\frac{3}{2}-\nu -\kappa+ \frac{3}{2}\epsilon},
\ee
where
\be \label{Akappa}
A_k= -\frac{\epsilon}{\sqrt{2}}\ e^{i\left(\nu-\frac{1}{2}\right)\frac{\pi}{2}}\   2^{\nu-\frac{3}{2}} \frac{\Gamma(\nu)}{\Gamma(3/2)} \frac{(1+ \epsilon)H_I}{\kappa c}\     k^{-\nu}
\ee  %il /\sqrt{2} è dovuto al termine in k al denominatore
does not depend on time, but only on the momentum $k$. Accordingly, the differential equation \eqref{betapert} can be solved analytically. The corresponding solution of Eq. \eqref{betapert} is given by
\begin{widetext}
\be \label{beta}
\beta({\bf x}, \tau) =  \left(  -\frac{2\ A_k\  \left(-\tau \right)^{\frac{7}{2}+\frac{3}{2}\epsilon-\kappa-\nu}}{\left( 3+\epsilon-2\kappa-2 \nu \right) \left( 7/2+3/2 \epsilon-\kappa-\nu \right)} +  \frac{\left(-\tau\right)^{2+\epsilon}}{2+\epsilon} c_1+c_2\right)  e^{i {\bf k}\cdot {\bf x}}.
\ee
\end{widetext}

From Eq. \eqref{S6} we notice that $h_{ij}^\parallel({\bf x}, \tau) \propto k_i k_j \beta({\bf x}, \tau)$ and since we are dealing with super-Hubble scales with $k \ll 1$ (see Fig. \ref{fig3}), this contribution can be neglected with respect to $h({\bf x}, \tau)$, which is given by\footnote{It is sufficient to subtract Eq. \eqref{S9} from Eq. \eqref{S8} recalling that $\Phi \equiv \Psi$ and neglect the gradient term, which again goes $\propto k^2\ \beta$.}
\begin{widetext}
\begin{align} \label{perth}
h({\bf x}, \tau)&\equiv h_k(\tau) e^{i {\bf k} \cdot {\bf x}} =-6 \beta^{\prime \prime}({\bf x}, \tau)-12 \frac{a^\prime}{a} \beta^{\prime}({\bf x}, \tau)= -6 \beta^{\prime \prime}({\bf x}, \tau)+ \frac{12(1+\epsilon)}{\tau} \beta^{\prime}({\bf x}, \tau) \notag \\[6pt]
& = \left[ \frac{\left( 6-6\epsilon-12 \kappa-12 \nu \right) A_k}{3+\epsilon-2\kappa-2\nu}\ \left( -\tau \right)^{\frac{3}{2}-\nu-\kappa+\frac{3}{2}\epsilon}+ 6 c_1 (1+\epsilon) \left(-\tau \right)^\epsilon \right] e^{i {\bf k} \cdot {\bf x}}.
\end{align}
\end{widetext}
We see that the value of $c_2$ does not affect $h({\bf x}, \tau)$, which is the physical perturbation. For this reason, we can safely set $c_2=0$. The constant $c_1$ is in principle arbitrary. However, it can be fixed by imposing that the total number of particles produced is a gauge-invariant quantity, i.e., exploiting the results of Sec. \ref{sez4}.

The perturbation tensor in synchronous gauge then reads
\be \label{eq36}
h_{\mu \nu}^S= \begin{pmatrix} 0 & 0 & 0 & 0\\
0 & h/3 & 0 & 0 \\
0 & 0 & h/3 &0 \\
0 & 0 & 0 & h/3
\end{pmatrix},
\ee
where again we remark we are dealing with super-Hubble scales and $h$ is given by Eq. \eqref{perth}.

\end{document}